\documentclass[%
 reprint,
groupedaddress,
showpacs,
 amsmath,amssymb,
 aps,
prstab,
floatfix,
]{revtex4-1}

\usepackage{graphicx}
\usepackage{dcolumn}
\usepackage{bm}
\usepackage{siunitx}
\usepackage{xcolor}

\begin{document}


\title{Simultaneous beam based alignment measurement for multiple magnets }

\author{Xiaobiao Huang}%
\email{xiahuang@slac.stanford.edu}
\affiliation{
SLAC National Accelerator Laboratory, Menlo Park, CA, 94025
}

\date{\today}

\begin{abstract}
We propose a method to simultaneously determine the magnetic centers of multiple 
quadrupoles in a transport line or a storage ring. 
The method finds the magnet centers by correcting the orbit shift due to a change of the quadrupole gradient strengths with orbit correctors. 
The quadrupoles are selected with orbit corrector magnets and beam position monitors in between
to ensure  that orbit correction at the quadrupole locations can be achieved. 
The correction of the induced orbit shift is done by steering the orbit toward the quadrupole centers with 
correctors, using its response matrix with respect to the correctors. 
The response matrix can be measured or calculated. 
Simulations with a section of the Linac Coherent Light Source (LCLS) II and the SPEAR3 storage ring 
are done to demonstrate the feasibility and performance of the method. 
It is also experimentally tested on SPEAR3. 
The method can be extended for beam based alignment measurement of nonlinear magnets. 
\end{abstract}

\maketitle

\section{Introduction}
Despite the ever improving survey and positioning technology, misalignment of magnets in accelerators is inevitable. 
Misalignment causes beam orbit offsets from the centers of the quadrupole and nonlinear (i.e., sextupole and octupole) magnets. 
In storage rings, the ``feed-down'' effects of the multipole magnets introduce linear optics errors, 
coupling errors, chromatic errors, and degradation to nonlinear beam dynamics performances. 
In linacs, the orbit offsets in quadrupoles cause dispersive errors, which leads to emittance dilution. 
In addition, such orbit offsets complicate the tuning of the quadrupoles as any change of gradient will lead to 
downstream trajectory shifts. 
Finding the magnetic centers with beam-based methods and steering the beam through the magnetic centers of the magnets 
have many benefits. 
Beam based alignment (BBA) for quadrupole magnets has become a standard practice at modern accelerator facilities. 

BBA can be done with a model dependent approach or 
a model independent approach. 
In the model dependent approach, the orbit shift due to a change of the quadrupole gradient is measured and, by the use of 
a lattice model, the corresponding kick angle at the quadrupole location is calculated, from which the orbit offset is obtained~\cite{ROJSEL1994374,BrinkmannEPAC94,EndoBBA96}. 
The variation of the quadrupole gradient can be done through a low frequency harmonic modulation, which leads to 
an orbit modulation of the same frequency~\cite{Barnett95}. The harmonic modulation reduces noise effects  and 
improves the measurement accuracy. 

In the model independent approach, the goal is to find an orbit through the quadrupole on which a change of the quadrupole strength 
does not cause a deflection of the beam orbit. This can be achieved by experimentally steering the orbit with 
a corrector magnet, while observing the orbit shift by the quadrupole variation at each step. 
This could be done manually~\cite{Rice83NS}. 
A commonly used method is implemented in the Matlab Middle Layer~\cite{PortmannMML}, for which  
the  quadrupole center offset is found by interpolating the orbit shifts due to the quadrupole 
gradient variation with respect to the beam orbit to find the zero-crossing~\cite{PortmannBBA}. 
The linear curves of the orbit shift at many locations vs. the beam orbit at a beam position monitor (BPM) adjacent to the quadrupole 
makes a ``bow-tie'' plot, on which the quadrupole center can be easily recognized. 
The model independent method does not require an accurate lattice model and 
can find the BPM reading corresponding to the quadrupole center on the adjacent BPM. 
BPM calibration errors and electrical offsets have no negative impact on the results. 

A recent progress on the topic is the use of AC excitation of corrector magnets for  beam based alignment~\cite{MartiACBBA}. 
The orbit shifts at two selected BPMs are linearly related and the slope of dependence will change when the quadrupole strength is 
varied. The intersection of the two linear curves, with or without quadrupole strength variation, gives the position of the 
quadrupole center. This method is fast because beam orbit measurement with AC excitation is fast. In addition, horizontal 
and vertical orbit excitation can be done simultaneously with different driving frequencies. 
For BBA of quadrupoles in storage rings, typically only one magnet is changed at a time. 

Reference~\cite{TenenbaumBBA} discusses a few techniques for beam-based alignment for linacs~\cite{LavineLinac88, AdolphsenPAC89, EMMA1999407, RAUBENHEIMER1991191}. These methods are similar to the methods employed in rings in measuring the trajectory shifts due to 
a variation of the quadrupole gradients, although in this case, the variation can be introduced by turning off the selected quadrupoles 
or measuring the trajectory differences of the electron and positron beams (for linear colliders). 
Most of these methods are model dependent as they solve the quadrupole offsets and BPM offsets from the measured trajectory 
shifts with the use of transfer matrices computed with a model~\cite{LavineLinac88, AdolphsenPAC89, EMMA1999407}. 
However, in one method, the goal is to correct the beam trajectory and simultaneously the 
trajectory shifts due to the scaling of the strengths of all quadrupoles~\cite{RAUBENHEIMER1991191}. 
This method, referred to as dispersion free (DS) correction, does not aim at finding the offsets of the individual quadrupole 
magnets, but the minimization of the combined effect of the quadrupole misalignment to beams with energy errors. 

Reference~\cite{TalmanPAC03} proposes a BBA method for  quadrupole families on serial power supply. 
The key idea is to restore the orbit after the modulation of quadrupole strengths with correctors on or next to the quadrupoles and to deduce the initial orbit offsets from the change of corrector strengths. 
This method was later tested in experiments~\cite{PINAYEV2007351}.
A BBA method to address the challenging situation in the interaction region of colliders is 
discussed in Reference~\cite{HoffstaetterPRAB02}. 

In this paper, we propose a beam-based method to find the quadrupole magnetic centers for multiple magnets simultaneously. 
This is achieved by correcting the orbit shifts due to variations of the quadrupole gradients, while the group of 
quadrupoles are selected to make the correction possible and easy to do. The method is applicable to both linacs and storage rings. 
The proposed method is similar to the DS method in correcting the orbit shift induced by quadrupole gradient variations. 
However, in our case, the goal is to determine and register the quadrupole center offsets with BPMs. 
Therefore, the resulting orbit offsets after the correction are not an issue. 
This is a model independent BBA method, as the quadrupole offsets found by the method does not 
require or depend on a lattice model, even though such a model could be used to calculate the response matrix~(which could also be measured),
and are not affected by BPM calibration errors or electrical offsets. 
The pattern of gradient changes can be properly chosen to facilitate the measurements, for example, by alternating the signs  
of gradient variations to keep a stable beam in ring applications. 
This method could be extended for nonlinear magnets in storage rings. 

This method is also similar to the method discussed in Reference~\cite{TalmanPAC03} in that both methods use correctors to determine the centers of multiple quadrupoles simultaneously.
However, there are several key differences between the two. 
First, the proposed method use correctors to alter the orbit at the quadrupole locations such that the induced orbit drift is set to zero (or minimized, in practice), while in Reference~\cite{TalmanPAC03} the method aims at restoring the orbit to before the quadrupole modulation is applied. 
Second, our method registers the quadrupole centers directly with nearby BPMs, while the method in Reference~\cite{TalmanPAC03} uses the changes of strengths of the nearby correctors to deduce the orbit offsets at the quadrupole. 
Corrector at or near the quadrupoles are required for the latter, which cannot always be satisfied, while the proposed method requires only enough correctors to independently change the orbit at the quadrupoles in the group. 

The main benefit of the proposed method is to substantially expedite BBA  by parallelizing the process. 
We may refer to the method as parallel BBA (P-BBA).
The method could have a crucial impact to the 
commissioning of new accelerators. It will also enable more frequent routine BBA measurements on operating machines. 

The paper is organized as follows: Section~\ref{sec:Linac} discusses the method for applications to linacs, including 
detailed descriptions of the theory and simulations for a section of the Linac Coherent Light Source (LCLS) II~\cite{LCLSII}; 
Section~\ref{sec:Ring} discusses the method for storage rings and demonstrates it with the application to the 
SPEAR3 storage ring~\cite{spear3} in both simulation and experiments; 
Section~\ref{sec:Nonlinear} briefly discusses the special considerations for applying the method to nonlinear magnets; 
and Section~\ref{sec:Conclu} gives the conclusions.

\section{P-BBA for a transport line \label{sec:Linac}}
\subsection{The method \label{sec:BBAmethodLinac}}
\begin{figure*}[htb]  
\centering
\includegraphics[width=4.in]{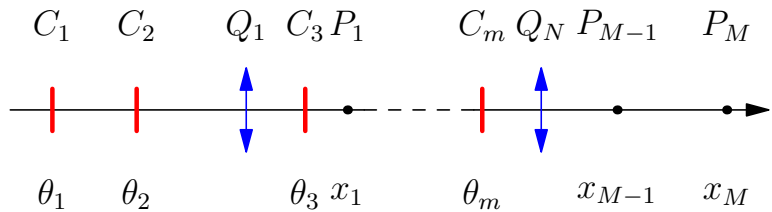}
\caption{ Schematic of an accelerator section for parallel BBA.
}
\label{fig:schematic}
\end{figure*}

In the following we consider BBA for quadrupoles in a 
transport line. 
Figure~\ref{fig:schematic} is a schematic of the lattice section, which 
consists of quadrupole magnets, orbit correctors, and BPMs. 
The magnetic centers of the quadrupoles are at $\Delta_i$ and 
the beam trajectory passes through the quadrupoles with position coordinate $\bar{x}_{i}$, for 
$i=1$, 2, $\cdots$, $N$. 
The  quadrupole centers relative to the beam orbit are 
$\bar{X}_i=\Delta_i-\bar{x}_i$.
The quadrupoles can be modeled as thin-lens elements. 
For quadrupole $i$, the nominal integrated gradient is labeled $K_{i0}$, while the change is labeled $k_i$, and the strengths after 
changes is $K_i=K_{i0}+k_i$. 

The beam receives an angular kick by the quadrupole when the trajectory is off-centered. 
When the quadrupole strength is changed, the kick angle  will also change. 
The kick angle change by quadrupole $i$ will be
\begin{align}\label{eq:deltaPhikick0}
    \Delta \phi_i = k_i(\Delta_i-\bar{x}_i-\Delta \bar{x}_i(\Delta \phi_1,\Delta \phi_2,\cdots,\Delta \phi_{i-1})),
\end{align}
where $\Delta \bar{x}_i(\cdot)$ is the trajectory shift at quadrupole $i$ due to the kick angle changes by 
the upstream quadrupoles. The $\Delta \bar{x}_i(\cdot)$ term in the kick angle change is  nonlinear 
with respect to quadrupole strength changes as the effects of upstream quadrupoles cascade upon each other. 
However, we can choose the size of $k_i$ to make the nonlinear terms small. 
And, as we steer the beam through the centers of the quadrupoles, the kick angle changes will diminish 
and in turn so do these nonlinear terms. 
In the following we drop the $\Delta \bar{x}_i(\cdot)$ terms, and hence
\begin{align}
    \Delta \phi_i = k_i(\Delta_i-\bar{x}_i). 
\end{align}

The kick angle changes due to gradient variations will cause the beam trajectory to change. 
We refer to such changes as the induced trajectory (or orbit) shifts. 
The induced trajectory shift at BPM  $P_i$ is the position 
component of
\begin{align}
    {\bm \xi}^{(i)} = \sum_{j=1}^{Q<P_i} {\bf M}(P_i|Q_j)\begin{pmatrix} 0 \\ \Delta \phi_j
    \end{pmatrix} , 
\end{align}
where ${\bf M}(P_i|Q_j)$ is the transfer matrix from quadrupole $Q_j$ to BPM $P_i$ and 
the condition $Q<P_i$ indicates quadrupoles upstream of $P_i$. 
If we label  the (1,2) element of ${\bf M}(P_i|Q_j)$ as ${ A}_{12}^{(ij)}$, 
the trajectory shift at BPMs can be written as 
\begin{align}\label{eq:xp0}
    \xi_i=\sum_{j=1}^{Q<P_i} { A}_{12}^{(ij)} k_j (\Delta_j-{\bar x}_j),
\end{align}
which can be written in a matrix form as
\begin{align}\label{eq:xp0Mat}
    {\bm \xi}= 
    {\bf A} {\bf k}  ({\bm \Delta}-{\bar x}),
\end{align}
where ${\bf k}$ is a diagonal matrix whose ($j$,$j$) element is $k_j$,
${\bf A}$ is a matrix of dimension $M\times N$ with its ($i$,$j$) element being 
${ A}_{12}^{(ij)}$ and zero if quadrupole $Q_j$ is downstream of BPM $P_i$, 
and ${\bm \Delta}$ and ${\bar x}$ are vectors formed with $\Delta_j$ and $\bar{x}_j$, $j=1$, 2, 
$\cdots$, $N$, respectively. 

Orbit correctors can change the trajectory at the quadrupole locations. 
The changes can be calculated using transfer matrices from the correctors to the quadrupoles. 
At quadrupole $Q_j$, the trajectory will be the position component of 
\begin{align}
    \bar{\bf x}^{(j)} = \bar{\bf x}^{(j)}_0 +  \sum_{l=1}^{C<Q_j} {\bf M}(Q_j|C_l)\begin{pmatrix} 0 \\ \ \theta_l
    \end{pmatrix}, 
\end{align}
where $\bar{\bf x}^{(j)}_0$ is the coordinates at quadrupole $Q_j$ when the correctors are at the initial 
values (i.e., $\theta_l=0$ for $l=1$, 2, $\cdots$, $m$),  
${\bf M}(Q_j|C_l)$ is the transfer matrix from corrector $C_l$ to quadrupole $Q_j$ and 
the condition $C<Q_j$ represents correctors before the quadrupole. 
The trajectory at all quadrupoles can be written in the matrix form as 
\begin{align}\label{eq:xQmat}
    \bar{\bf x}({\bm \theta}) = \bar{\bf x}_0 + {\bf C}{\bm \theta}, 
\end{align}
where $\bar{\bf x}$ is a $N$-dimensional vector with its component being the position coordinates 
at the quadrupoles, $\bar{\bf x}_0=\bar{\bf x}({\bm 0})$, ${\bf C}$ a $N\times m$ matrix whose 
($j$,$l$) element is the (1, 2) element of ${\bf M}(Q_j|C_l)$  if $C_l$ is upstream of $Q_j$ or 
zero otherwise, and ${\bm \theta}$ is an $m$-dimensional vector with all the corrector kick angles as its 
elements. 

Combining Eqs.~\eqref{eq:xp0Mat} and \eqref{eq:xQmat}, we obtain a relationship between the induced 
trajectory shift by the quadrupole gradient changes and the kick angles of the correctors,
\begin{align} \label{eq:xiontheta}
    {\bm \xi}&=  {\bf A} {\bf k}  ({\bm \Delta}- \bar{\bf x}_0 - {\bf C}{\bm \theta}), \\
    &= {\bm \xi}_0 + {\bf R}{\bm \theta},  \label{eq:xiontheta2}
\end{align}
where ${\bm \xi}_0={\bf A} {\bf k}  ({\bm \Delta}- \bar{\bf x}_0 ) $ is the induced trajectory shift when  ${\bm \theta}=0$ 
and 
\begin{align}
    {\bf R} \equiv \frac{\partial {\bm \xi}}{\partial {\bm \theta}}=  -{\bf A} {\bf k} {\bf C}
\end{align}
is the response matrix of the induced trajectory shift with respect to the corrector kick 
angles. 

The goal of the P-BBA method is to find corrector kick angles, ${\bm \theta}$, to set the induced 
trajectory shift to zero. 
Knowing the response matrix, ${\bf R}$, and the measured induced trajectory shift, ${\bm \xi}_0$, 
the changes to the corrector kick angles required to eliminate the induced trajectory shift are given by
\begin{align}\label{eq:deltatheta}
    {\bm \theta} =  
    -({\bf R}^T{\bf R})^{-1}{\bf R}^T {\bm \xi}_0. 
\end{align}
Because the induced trajectory shift is measured at multiple BPMs 
and all measurements have errors, in reality the goal will not be 
achieved exactly. Instead, we aim at minimizing the induced trajectory shift through 
a least-square problem, i.e., we minimize
\begin{align}
    \chi^2 = {\bm \xi}^T{\bm \xi}. 
\end{align}
This can be achieved iteratively. At each iteration, Eq.~\eqref{eq:deltatheta} can be used to 
calculate the required changes to the kick angles toward the next step. 

For the scheme to work, the matrix inversion in Eq.~\eqref{eq:deltatheta} needs to have a unique 
solution. In other words, the quadrupoles, correctors, and BPMs should be chosen to avoid degeneracy in 
matrices ${\bf A}$ and ${\bf C}$ (the diagonal matrix ${\bf k}_m$ will be non-degenerate as the quadrupole 
gradients are changed). 
The ${\bf A}$ matrix will be non-degenerate if no two kick patterns by the selected quadrupoles cause the same 
trajectory shift on the BPMs. This requires at least two BPMs downstream of the last quadrupole 
and, for any two consecutive quadrupoles, there is either at least one BPM in between, or two BPMs in the 
space before the next quadrupole. 
The ${\bf C}$ matrix will be non-degenerate if the correctors can steer the beam to the desired trajectory
at all quadrupoles. 
This requires at least one corrector upstream of the first quadrupole and, for  any two consecutive quadrupoles,
there are a pair of 
correctors in the space upstream or at least one corrector in between. 

It is preferable to use all correctors and BPMs available as it helps increase the level of correction 
precision. 
Therefore, we only need to select the group of quadrupoles for simultaneous BBA measurements. 
Usually we can divide all quadrupoles in a beamline into several groups, each group consisting of
quadrupoles with a large distance in between, possibly with some quadrupoles skipped. 
For example, the first, fourth, seventh, $\cdots$, quadrupoles can be put in one group; 
the second, fifth, eighth, $\cdots$ in another group, etc. 
For a long beamline with many quadrupoles, it may be necessary to first divide it into several sections and group 
the quadrupoles in each section as described in the above. 
This is because of the cascading effects of the induced trajectory shift due to upstream quadrupoles 
at downstream quadrupoles (see the $\Delta\bar{x}$ term in Eq.~\eqref{eq:deltaPhikick0}). 
We would like the higher order effects to be much smaller than the direct effect.

The pattern of gradient changes, ${\bf k}$, can be a simple scaling change to the initial values, if no quadrupole 
involved is particularly weak. 
For example, all quadrupole power supplies can be reduced by $5\%$. 
A pattern with equal changes of integrated gradients but with alternating signs can also be used. 
It is worth noting that if multiple quadrupoles are on a serial power supply, their magnetic centers can 
still be resolved with the proposed method, as long as there are correctors and BPMs between these quadrupoles 
to detect and correct their individual contributions to the induced trajectory shift. 

\subsection{Error estimate}
Because we can select the target quadrupoles  according to the available 
corrector magnets and BPMs, we can correct the trajectory at the quadrupole locations to the accuracy 
of measurements for the induced trajectory shifts by the BPMs. 
The BPM measurement errors and the quadrupole center errors are related through 
Eq.~\eqref{eq:xiontheta}. 
If we define 
\begin{align}
    {\bf R}_Q\equiv \frac{\partial{\bm \xi}}{\partial{\bf \Delta}}={\bf Ak}
\end{align}
as the response matrix of the induced trajectory shift with respect to the quadrupole center offsets, 
the covariance matrix of the errors in the measured quadrupole offsets, ${\bm \Sigma}_{\Delta\Delta}$,
is related to the BPM measurement errors through
\begin{align}
    {\bm \Sigma}_{xx}\equiv \langle ({\bm \xi}-\bar{\bm \xi})({\bm \xi}-\bar{\bm \xi})^T \rangle
    ={\bf R}_Q{\bm \Sigma}_{\Delta\Delta}{\bf R}_Q^T, 
\end{align}
where $\langle \cdot \rangle$ represents ensemble average over many measurements and 
${\bm \Sigma}_{xx}$=$\text{diag}$($\sigma_1^2$, $\sigma_2^2$, $\cdots$, $\sigma_M^2$), 
with $\sigma_i$, $i=1$, 2, $\cdots$, $M$ being the error sigma of the BPMs. 
Therefore, 
\begin{align}
{\bm \Sigma}_{\Delta\Delta} = ({\bf R}_Q^T {\bf R}_Q)^{-1}{\bf R}_Q^T{\bm \Sigma}_{xx}{\bf R}_Q({\bf R}_Q^T {\bf R}_Q)^{-1}. 
\end{align}
If all BPMs have the same measurement error sigma, $\sigma_\text{BPM}$, we have
\begin{align}
{\bm \Sigma}_{\Delta\Delta} = \sigma^2_\text{BPM}({\bf R}_Q^T {\bf R}_Q)^{-1}. 
\end{align}
The diagonal elements in the ${\bm \Sigma}_{\Delta\Delta}$ matrix give the variance of the 
quadrupole center offset measurements. 

\subsection{Simulation}
Simulation has been done to test the proposed P-BBA method. The accelerator modeling code Accelerator Toolbox~\cite{AccelTool} is 
used for the simulation.
The soft X-ray linac-to-undulator (LTU) section of the LCLS-II copper linac~\cite{LCLSII} is used in the study. 
The number of relevant elements in the line section are listed in Table~\ref{tab:paraLTUS}, including 
two correctors upstream of the section for each plane and 5 BPMs downstream of the section. 

\begin{table}[htbp] 
\begin{center} 
\caption{Elements of the LCLS-II copper soft X-ray LTU section used in simulation. 
Two correctors in each plane upstream of the section and 5 BPMs downstream of the section 
are added to the system. 
\label{tab:paraLTUS} } 
\begin{tabular}{|c|c|} 
\hline
Parameter & Value \\
\hline 
Length (m) & 372.5 \\
number of quadrupoles &  33 \\
number of H correctors & 16+2 \\
number of V correctors & 17+2 \\
number of BPMs & 41+5 \\
\hline 
\end{tabular} 
\end{center} 
\end{table} 

Random misalignment errors are first added to the quadrupoles in the section, with rms offsets of 100 $\mu$m 
for both transverse planes. 
The quadrupole misalignment  causes a distorted beam trajectory (w/ zero initial launching angle and position). 
Corrector magnets are used to restore the trajectory toward the target ($x=y=0$ at all BPMs). 
The rms trajectory errors are corrected to below $20$~$\mu$m. 
The maximum kick angle by the correctors for the two planes is $33$~$\mu$rad (H) and  $61$~$\mu$rad (V), 
respectively. 

\begin{figure}[htb]  
\centering
\includegraphics[width=3.2in]{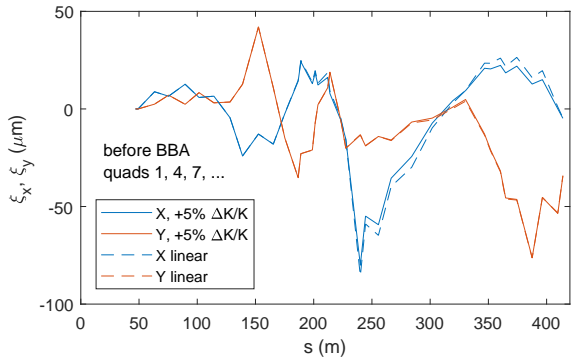}
\caption{ The induced trajectory shift by a $5\%$ increase of the strengths of group 1 quadrupoles. 
Solid lines show actual values obtained by tracking. The dashed lines (`linear') show the values 
scaled up from a $0.1\%$ gradient change by a factor of 50. 
}
\label{fig:inducedTrajShiftBefore}
\end{figure}

We divide the 33 quadrupoles in the LTU section into three groups according to their locations. 
Group 1 consists of quadrupoles 1, 4, 7, $\cdots$, 31; 
group 2 consists of quadrupoles, 2, 5, 8, $\cdots$, 32; 
and group 3 consists of quadrupoles, 3, 6, 9, $\cdots$, 33. 
Figure~\ref{fig:inducedTrajShiftBefore} shows the induced trajectory shift when the strengths of 
all quadrupoles in group 1
 are scaled up by $5\%$. 
The trajectory shift is up to 80~$\mu$m. 
Also shown in the figure are the induced trajectory shift by the linear model (obtained by scaling up 
the induced trajectory shift of a tiny gradient change). 
It can be seen that the higher order terms only cause a small deviation from the linear model at the 
downstream BPMs. 

\begin{figure}[htb]  
\centering
\includegraphics[width=3.2in]{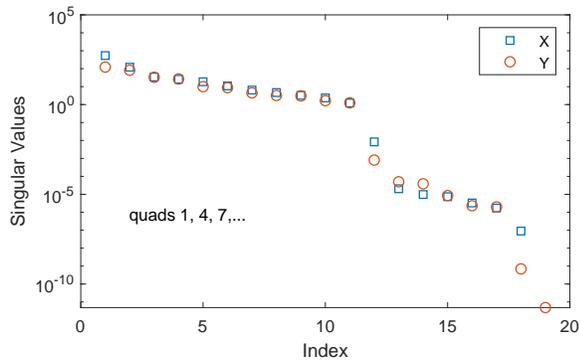}
\caption{ Singular values of the horizontal and vertical response matrices of 
the induced trajectory shift by a $5\%$ increase of the strengths of group 1 quadrupoles with respect to the 
corrector magnets. 
}
\label{fig:SVrespmatIOSgrp1}
\end{figure}

The response matrix of the induced trajectory shift with respect to the correctors is calculated with the 
lattice model. Figure~\ref{fig:SVrespmatIOSgrp1} shows the singular values of the response matrices of the 
induced trajectory shifts with respect to the correctors for both transverse planes for the 
group 1 quadrupoles. While the dimensions of the response matrices are $46\times18$ and 
$46\times19$, respectively, for the horizontal and vertical planes, there are only 11  modes with 
substantial singular values. This is because there are only 11 quadrupoles. 
The other SV modes would be exactly zero if not for the higher order effects. 

\begin{figure}[htb]  
\centering
\includegraphics[width=3.2in]{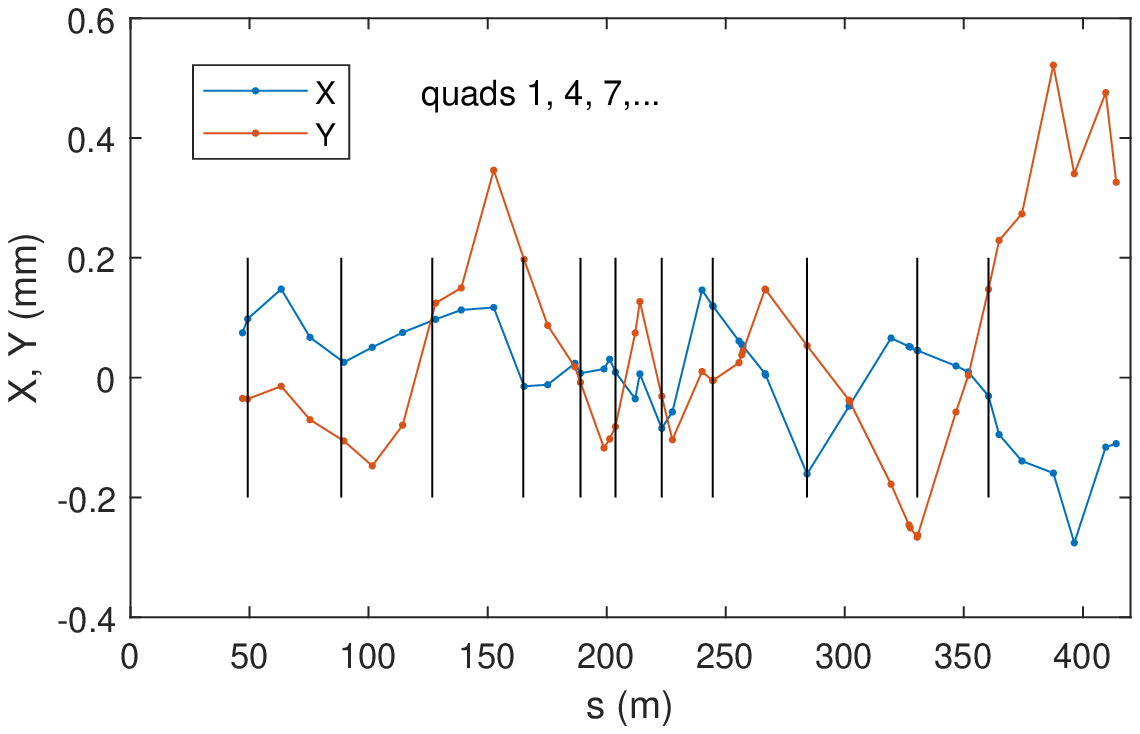}
\includegraphics[width=3.2in]{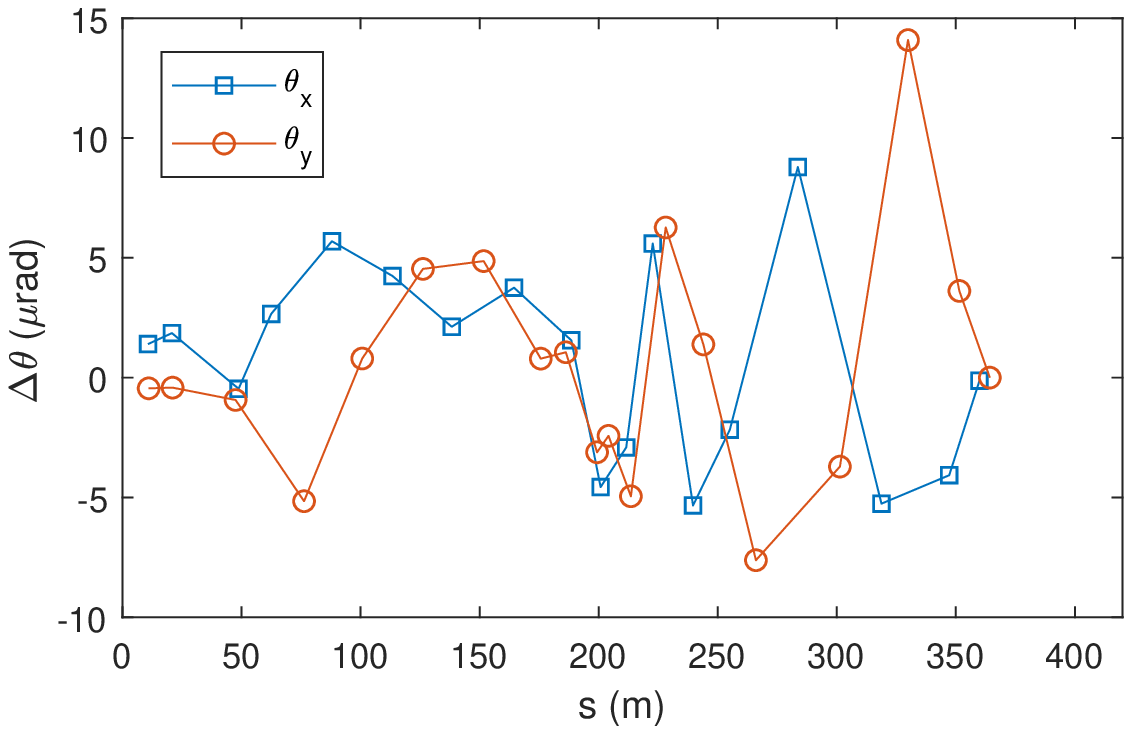}
\caption{ Top: the beam trajectory after correction of the induced trajectory shift. Quadrupole locations in the 
group are marked with vertical lines. 
Bottom: the required kick angle changes on the correctors for the correction. 
}
\label{fig:XYdthetaGrp1}
\end{figure}

The correction of the induced trajectory shift is done with Eq.~\eqref{eq:deltatheta}, using only 
the 11 leading singular values in the matrix inversion calculation. 
With one iteration, the rms values of the induced trajectory shift on the 46 BPMs are reduced to 
$0.35$~$\mu$m and $0.02$~$\mu$m 
for the horizontal and vertical planes, respectively, when no measurement errors are included to the 
BPMs. A second iteration reduce them further to 5~nm and 
$0.1$~nm, respectively. 
The required kick angle changes for the correction are mostly below 10~$\mu$rad. 
Figure~\ref{fig:XYdthetaGrp1} shows the beam trajectory after the correction of the induced trajectory shift 
and the required changes to the corrector kick angles. 

In simulation, the quadrupole center offsets can be found with the corrected lattice 
by tracking a particle with initial coordinates of 
$x=0$ and $x'=0$ to the quadrupole locations. 
The differences with the target values are on the a few nano-meter level. 
Essentially, the quadrupole centers can be exactly found if there is no BPM measurement errors.

\begin{figure}[htb]  
\centering
\includegraphics[width=3.2in]{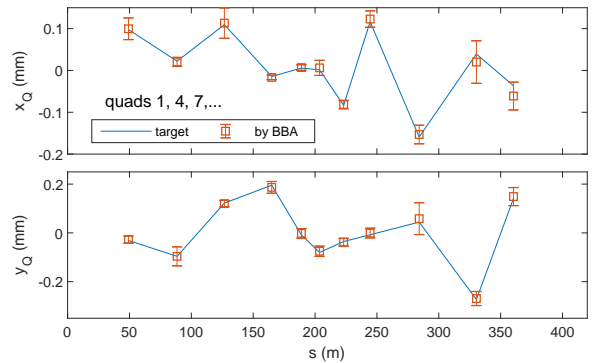} 
\caption{ The quadrupole offsets found with the P-BBA method, with BPM error sigma at 5~$\mu$m. 
}
\label{fig:offsetcmpGrp1}
\end{figure}

With the BPM error sigmas set to $\sigma_\text{BPM}=5$~$\mu$m, the correction method is repeated 10 times, 
from which the error bars to the quadrupole offsets can be found. 
Figure~\ref{fig:offsetcmpGrp1} shows the comparison of the measured quadrupole center offsets to the target values. 
The mean error sigmas for quadrupole offsets are 22 and 27~$\mu$m, for the horizontal and vertical places, respectively. 
The errors can be reduced by averaging in the measurement of induced trajectory shifts or increasing the 
gradient changes of quadrupoles. 

\begin{figure}[htb]  
\centering
\includegraphics[width=3.2in]{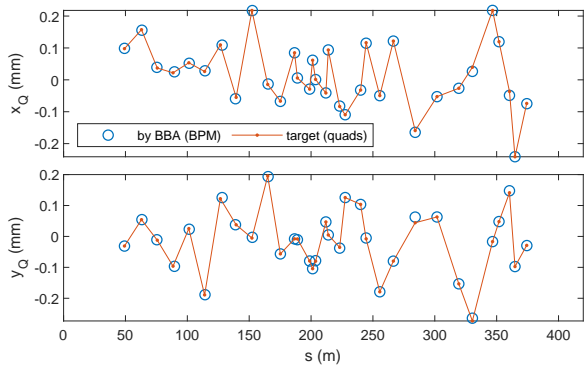} 
\caption{ The quadrupole center offsets registered by BPMs through the P-BBA method are compared to the target values 
at the quadrupole locations, for all quadrupoles in the LTU section. 
}
\label{fig:offsetcmpAll}
\end{figure}

On the real machine, we cannot determine the quadrupole centers by tracking particles to the quadrupole locations. 
Instead, we will use the readings of the BPMs near the quadrupoles in the group to represent the center positions of these 
quadrupoles. 
For the LTU section, every quadrupole is next to a BPM and hence the quadrupole center offsets can be accurately 
recorded. 
Simulation with quadrupoles in group 2 and group 3 yields similar results. 
Combining the results from all three groups, the  center offsets of all quadrupoles in 
the LTU section are found. 
Figure~\ref{fig:offsetcmpAll} shows the comparison of the BPM offset values found by the P-BBA method 
and the target offset values at the quadrupoles. 

Performing P-BBA for all 33 quadrupoles in the LTU section requires
only three times of correction of the induced trajectory shifts. Each time the quadrupole 
gradients are varied two to three times. 
The total time would be substantially less than the current method of making the ``bow tie'' plot for each individual 
quadrupole. 

\section{P-BBA for a storage ring \label{sec:Ring}}

\subsection{Method}
The method of performing simultaneous BBA for multiple quadrupoles by correcting the induced orbit shift (IOS)
can be applied to storage rings. 
Similarly, we select a group of quadrupoles that are sufficiently separated, with BPMs and correctors in between, 
and measure the IOS by varying the gradients of these quadrupoles. 
Corrector magnets are used to correct the IOS observed by the BPMs. Essentially, we are correcting the orbit 
at the locations of the selected quadrupoles toward the magnetic centers. 

The description of the method presented in section~\ref{sec:BBAmethodLinac} still largely applies, except now the BPMs 
measure the closed orbit, instead of the one-pass trajectory. 
The elements in the ${\bf A}$ matrix are now the orbit responses at the BPMs by the kicks at the quadrupole locations, 
while the elements in the ${\bf C}$ matrix are the orbit responses at the quadrupole locations by the corrector magnets. 
Since in a storage ring, an angular kick at any location affects the closed orbit everywhere, the ${\bf A}$ and ${\bf C}$ 
matrices are now full matrices. 

Simultaneous changes of the gradients of many quadrupoles can substantially change the linear optics of the ring, which could 
cause significant degradation of beam lifetime or move the beam across resonance conditions and in turn cause beam losses. 
Therefore, we should choose the quadrupoles carefully and apply a gradient change pattern, ${\bf k}$, properly, to ensure the 
beam will be stable during and after the gradient changes. 
For example, the signs of the gradient changes can be alternated in a sequence of quadrupoles to keep the betatron tunes 
nearly fixed. 
The number of the quadrupoles in a group can be limited to allow a relatively large gradient change while keeping the beam 
stable. 

\subsection{Simulation}
Simulations with the SPEAR3 storage ring are done to demonstrate the application of the P-BBA method to storage rings. 
SPEAR3 is a 3-GeV third generation synchrotron light source with a circumference of 234 meters~\cite{spear3}. 
The lattice consists of 18 double bend achromat (DBA) cells in a racetrack configuration, with 14 standard cells forming 
two arcs and 4 matching cells that flank the two long straight sections. 
There are a total of 97 quadrupole magnets in the lattice. 
There are 58 horizontal correctors and 56 vertical correctors. Currently 56 BPMs are used for beam orbit control. 

In the simulation we first introduce random misalignment errors to all quadrupoles with an rms offset of 200~$\mu$m 
for both planes. The orbit is then corrected to below 2~$\mu$m at the BPMs with the correctors. 
We select a group of 14 quadrupoles for simultaneous BBA as an example. These are the second QF magnet in each of the 14 
standard cells. The QF magnets are 0.35~m long and the nominal gradients are  about $1.9$~m$^{-2}$. 
We choose to alternate the gradient changes with a $+4\%$ change for the odd number quadrupoles and 
a $-4\%$ change for the other quadrupoles. 
The betatron tunes become to $\nu_x=14.070$ and $\nu_y=6.153$, down from the original values of 
 $\nu_x=14.106$ and $\nu_y=6.177$. 
 
\begin{figure}[htb]  
\centering
\includegraphics[width=3.2in]{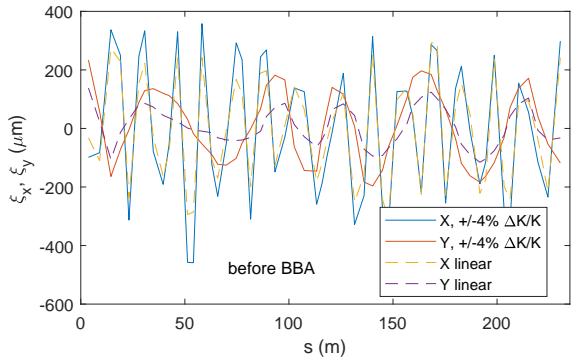} 
\caption{ The IOS in the SPEAR3 ring when the gradients of the selected 14 QF quadrupoles are 
varied by  $+4\%$ or  $-4\%$. 
The dashed lines are obtained by varying the gradients by only $+0.04\%$ or  $-0.04\%$ and scaling up the orbit shift linearly. 
}
\label{fig:initIOSxyQF2}
\end{figure}

The initial IOS by the  gradient changes of the 14 QF magnets is shown in Figure~\ref{fig:initIOSxyQF2}. 
Also shown in the plots are the expected orbit shift for a linear model (with respect to the quadrupole gradients), 
which is obtained by scaling up the response of a tiny gradient change by the same pattern. 
The differences between the actual orbit shift and the linear model reflect the changes to the linear optics of the ring 
when the quadrupoles are changed. 

\begin{figure}[htb]  
\centering
\includegraphics[width=3.2in]{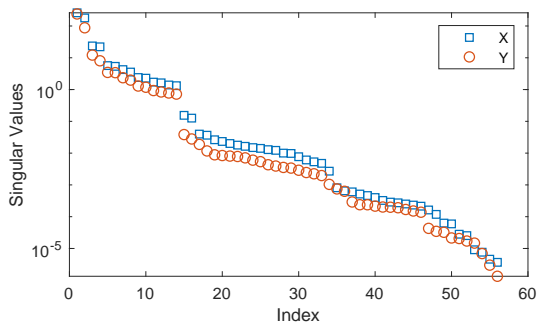} 
\caption{ The singular values of the horizontal and vertical response matrices of the IOS by the 14 QF magnets 
(with a scale change of  $+4\%$ or  $-4\%$ of alternating signs) 
with respect to 
the correctors. 
}
\label{fig:SP3SVQF2}
\end{figure}

\begin{figure}[htb]  
\centering
\includegraphics[width=3.2in]{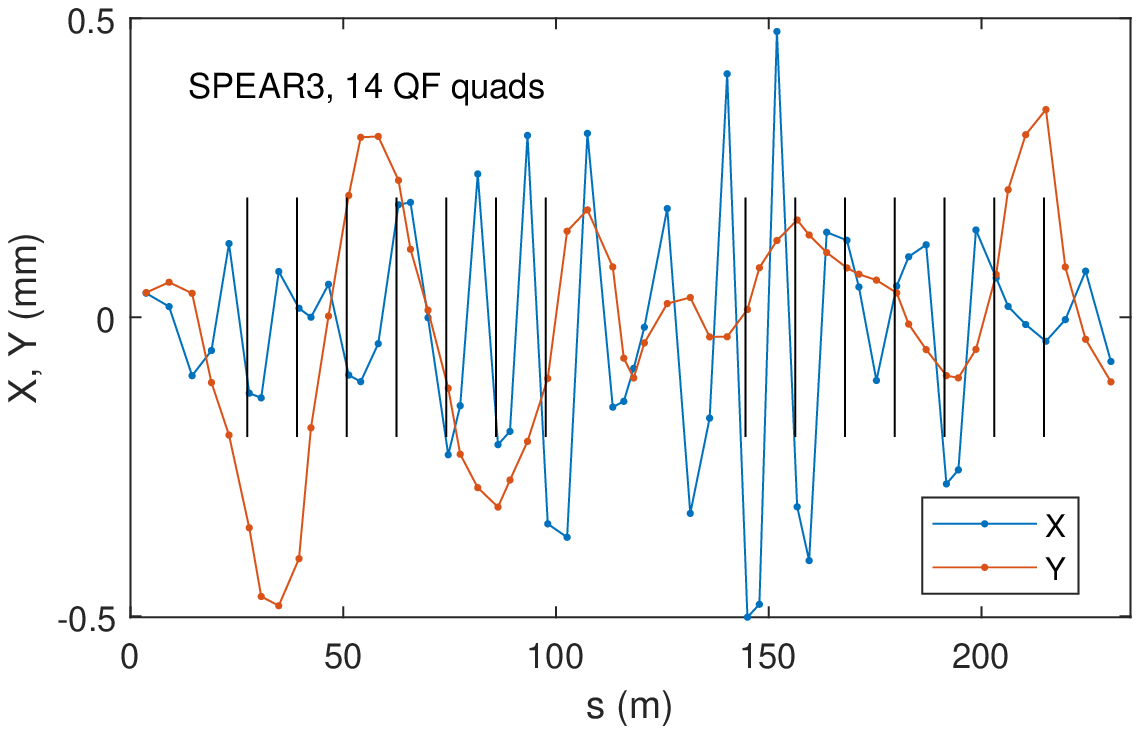} 
\includegraphics[width=3.2in]{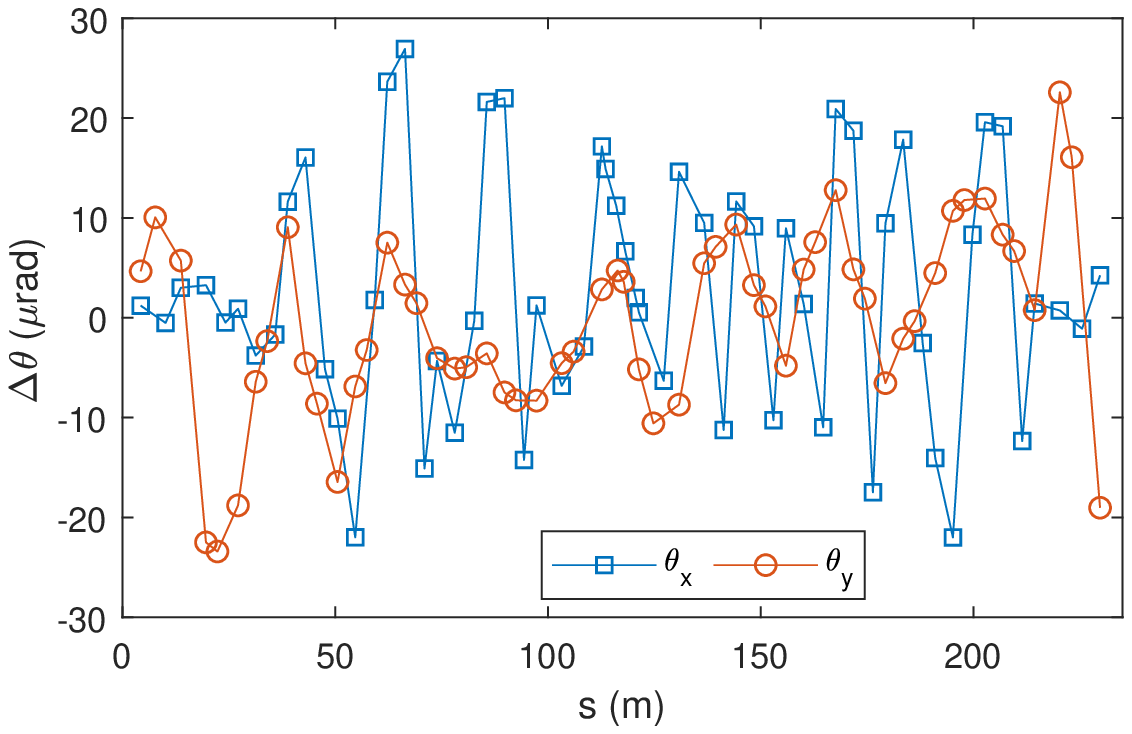} 
\caption{Top: the beam orbit after correction of the IOS for the SPEAR3 example. 
The locations of the quadrupoles in the BBA group are marked with vertical lines. 
Bottom: the corresponding corrector kick angles. 
}
\label{fig:SP3OrbitThetaxyQF2}
\end{figure}

The response matrices of the IOS with respect to the correctors are calculated with the design lattice model. 
Figure~\ref{fig:SP3SVQF2} shows the singular values of the horizontal and vertical response matrices. 
There are only 14 modes with significant singular values as there are 14 quadrupoles that affect the IOS. 
The calculated response matrices are used for the correction of the IOS with Eq.~\eqref{eq:deltatheta}. 
After three iterations of correction, the residual IOS after correction are reduced to sub-micron 
level. The differences between the quadrupole offsets found by BBA and the target values are also on the sub-micron level, 
as shown in Table~\ref{tab:SP3rmsIter}. 
The beam orbit is changed to go through the  centers of the 14 QF quadrupoles that are varied (see Figure~\ref{fig:SP3OrbitThetaxyQF2}). 
The required corrector kick angles are below 30~$\mu$rad. 
\begin{table}[htbp] 
\begin{center} 
\caption{Standard deviations of residual IOS at BPMs and 
the differences between the BBA results and the actual quadrupole centers (BBA error) after each iteration for 
the SPEAR3 example. \label{tab:SP3rmsIter} } 
\begin{tabular}{|c|c|c|c|c|} 
\hline
iteration & IOS-X & IOS-Y & BBA error (X) & BBA error (Y) \\
  & $\mu$m & $\mu$m & $\mu$m & $\mu$m\\ 
\hline 
$ 1$ &$23.9$ &$16.5$ &$19.0$ &$15.7$\\ 
$ 2$ &$ 3.2$ &$ 2.3$ &$ 3.2$ &$ 2.0$\\ 
$ 3$ &$ 0.6$ &$ 0.3$ &$ 0.4$ &$ 0.3$\\ 
\hline 
\end{tabular} 
\end{center} 
\end{table} 


\begin{figure}[htb]  
\centering
\includegraphics[width=3.2in]{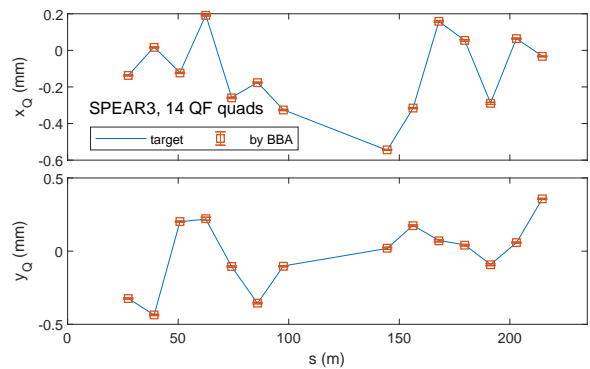} 
\caption{ The quadrupole center offsets found by the P-BBA method are compared to the target values for 
the SPEAR example. 
Error bars are by standard deviations of 10 random seeds, with BPM noise sigma of 1~$\mu$m. 
}
\label{fig:SP3OffsetcmpQF2}
\end{figure}

The BBA results are affected by BPM measurement errors. 
We repeated the P-BBA process 10 times, with random errors added to the orbit measurements and 
a BPM error sigma of  1~$\mu$m. 
The quadrupole center offsets are compared to the target values in Figure~\ref{fig:SP3OffsetcmpQF2}. 
The average error sigmas of the offsets are  $3.4$~$\mu$m (X) and  $5.4$~$\mu$m (Y) for the two transverse planes, respectively. 

\subsection{Experiments}
The P-BBA method has been experimentally tested on the SPEAR3 storage ring. 
In the experiment, the same 14 QF magnets as used in simulation  are targeted. 
The quadrupole gradients are changed by $\pm2\%$ in an alternating pattern. 

\begin{figure}[htb]  
\centering
\includegraphics[width=3.2in]{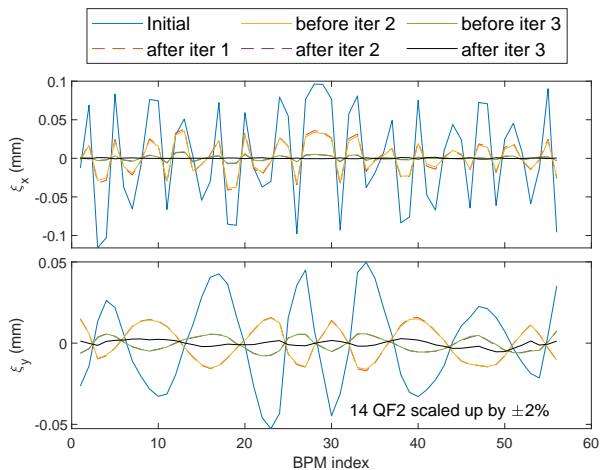} 
\caption{ The measured IOS by 14 QF magnets in a SPEAR3 experiment during 3 iterations of correction. 
The quadrupole gradient changes are $\pm2\%$ alternately. 
}
\label{fig:SP3QF2MeasIOS}
\end{figure}

Figure~\ref{fig:SP3QF2MeasIOS} shows the IOS measured during three iterations of correction. 
The initial IOS are up to 0.1~mm and 0.05~mm, respectively, in the horizontal and vertical planes. 
The conditions for `after iteration 1' and 'before iteration 2' are the same, so as 'after iteration 2' and 
'before iteration 3'. The measured IOS for these conditions overlap, which indicate the orbit shifts are 
reproducible. The rms IOS is reduced from $65.0$~$\mu$m to $0.6$~$\mu$m for the horizontal plane, 
and from $27.2$~$\mu$m to  $3.0$~$\mu$m in the vertical plane. 
It is noted that in each iteration it is an under-correction in the horizontal plane and an over-correction in 
the vertical plane, which could come from errors in the corrector strength calibrations. 
The over-correction on the vertical plane is $36\%$. 
The convergence would be much faster if we adjust the current to kick angle conversion coefficients for the correctors. 

\begin{figure}[htb]  
\centering
\includegraphics[width=3.2in]{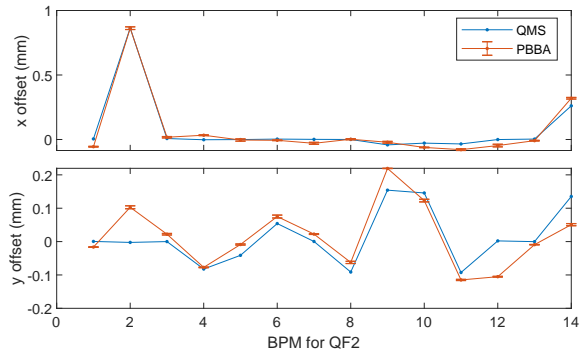} 
\caption{ The quadrupole center offsets from the initial beam orbit as measured by P-BBA in experiment are compared to 
the QMS offsets. 
}
\label{fig:SP3QF2MeasOffset}
\end{figure}

After the IOS correction, the quadrupole centers are 
marked by the BPMs next to the quadrupoles. 
The measurements are repeated 4 times with the same initial orbit, from which the error bars can be estimated. 
Figure~\ref{fig:SP3QF2MeasOffset} shows a comparison of the quadrupole center offsets  from the initial orbit 
measured by P-BBA and the conventional 'bowtie' method (quadrupole modulation system, or QMS)~\cite{PortmannBBA}. 
The initial orbit is different from the QMS offset orbit as steering is needed for injection or user beamlines. 
For example, the large horizontal offset at BPM 2 in the figure is to create a closed orbit bump at the injection septum. 
The P-BBA results are generally close to the QMS results. There are also some noticeable differences on the 
vertical plane, which would decrease if the vertical IOS correction is improved. 

One iteration of IOS correction with two quadrupole gradient modulation (one before, one after) took 32 seconds. 
A full correction with 3 iterations should take about 70 seconds if the two extra intermediate IOS measurements are 
skipped (14 second each). This is for the group of 14 quadrupoles. It would take less than 5 minute to find the offsets for all 
56 BPMs for SPEAR3. 
In comparison, it takes the current method (QMS) 2.5 hrs to complete BBA for the same quadrupoles.

\section{P-BBA for nonlinear magnets in storage rings\label{sec:Nonlinear}}
The BBA approach by correcting the IOS can be applied to sextupole and other nonlinear magnets in 
storage rings. The centers of several magnets can be found simultaneously, provided that varying their strengths 
does not cause beam loss and there are enough correctors and BPMs to correct the IOS. 

Large relative changes of strengths to the nonlinear magnets may be needed to induce large orbit shifts (in comparison 
to BPM errors). The number of nonlinear magnets that can be changed on such a scale while still keeping a stable 
beam may be limited. Groups of nonlinear magnets and special patterns of changes for them that are applicable for P-BBA 
could be found experimentally. 

The dependence of  IOS from variations of nonlinear magnets on the corrector magnets is not 
linear. Since the actual orbit offsets in the nonlinear magnets are not known, we cannot calculate the response matrix 
of the IOS with respect to the correctors with the lattice model. 
However, the response matrix can be measured on the machine for each iteration of the correction. 
To reduce the measurement time, it may be necessary to reduce the number of correctors used for the correction of 
induced orbit. 
For example, if we are trying to determine the center offsets of 20 sextupoles in a large ring with 300 correctors, 
there is no need to use all 300 correctors. Instead, it would be sufficient to choose 20 to 30 properly chosen corrector 
magnets. 
It may be possible to form combined orbit correction knobs with all correctors to target the orbits at the selected nonlinear 
magnets, using singular value decomposition on model calculated  orbit response matrix. 

Beam based optimization methods can also be used to find the orbit that minimizes the IOS. 
The Nelder-Mead simplex method~\cite{NelderMead1965} and the robust conjugate direction search (RCDS) method~\cite{RCDS} would 
be well suited for this application. 
Machine learning based optimization algorithms, such as the multi-generation Gaussian process optimizer~\cite{HuangMGGPO}, can also be used. 

\section{Conclusion\label{sec:Conclu}}
We proposed a method, P-BBA, to perform beam-based alignment measurements for multiple quadrupoles simultaneously. 
In the method, quadrupoles in the lattice are properly selected and grouped according to their locations relative to 
the corrector magnets and BPMs. The orbit shifts induced by a pattern of strength changes of the selected quadrupoles 
are measured with BPMs and corrected with the corrector magnets using the response matrix method with the aid of singular 
value decomposition. 
After the correction of the IOS, the beam orbit goes through the centers of the selected quadrupoles, subject only 
to BPM precision limitations. 
The method is applicable to one-pass systems such as linacs and transport lines, as well as storage rings. 

Simulations were done for a section of the LCLS-II and the SPEAR3 storage ring to demonstrate the method. In the 
LCLS-II example, quadrupole gradients are varied by $5\%$. In the SPEAR3 example, the gradients of the 
selected quadrupoles are varied by $+4\%$ or $-4\%$ in an alternating pattern to keep the betatron tunes nearly fixed. 
For both cases, the quadrupoles centers are found by the method and the  error sigmas for the quadrupole offsets are about 5 times of 
the BPM error sigma. 

The method was also experimentally tested on SPEAR3. We successfully demonstrated that the IOS are 
reproducible and can be corrected directly with orbit correctors, using model calculated response matrices. 
The offsets found by the P-BBA method generally agree with the conventional method. 
It is estimated that the P-BBA method is 30 times faster than the conventional method. 

Extension of the method to nonlinear magnets in storage rings is also discussed.

\section{Acknowledgements}
  This work was supported by the U.S. Department of Energy, Office of
  Science, Office of Basic Energy Sciences, under Contract No.
  DE-AC02-76SF00515

\end{document}